\documentclass[12pt]{article}
\usepackage{color}
\usepackage{amssymb,latexsym,bm, amsmath,amsthm, amsfonts,multirow, graphicx,mathrsfs}
\usepackage{eufrak,booktabs}
\usepackage{epsfig}
\graphicspath{{figs/}}

\def\Ab{{\boldsymbol A}}

\def\Eb{{\boldsymbol E}}

\def\Ib{{\boldsymbol I}}

\def\Kb{{\boldsymbol K}}

\def\Ob{{\boldsymbol O}}

\def\Qb{{\boldsymbol Q}}
\def\Rb{{\boldsymbol R}}
\def\Sb{{\boldsymbol S}}

\def\Ub{{\boldsymbol U}}
\def\Vb{{\boldsymbol V}}
\def\Wb{{\boldsymbol W}}
\def\Xb{{\boldsymbol X}}
\def\Yb{{\boldsymbol Y}}
\def\Zb{{\boldsymbol Z}}
\def\alphab{{\boldsymbol      \alpha}}
\def\betab{{\boldsymbol       \beta}}

\def\deltab{{\boldsymbol      \delta}}

\def\varepsilonb{{\boldsymbol \varepsilon}}
\def\zetab{{\boldsymbol       \zeta}}
\def\etab{{\boldsymbol        \eta}}

\def\mub{{\boldsymbol         \mu}}
\def\nub{{\boldsymbol         \nu}}

\def\yb{{\boldsymbol y}}
\def\zb{{\boldsymbol z}}

\def\Gammab{{\boldsymbol      \Gamma}}

\def\Sigmab{{\boldsymbol      \Sigma}}

\def\Omegab{{\boldsymbol      \Omega}}

\newcommand{\argmin}{\mathop{\rm argmin}}
\newcommand{\argmax}{\mathop{\rm argmax}}
\def\vec{{\rm vec}}

\def\1{{\bm 1}}
\def\0{{\bm 0}}

\def\cov{{\rm cov}}
\def\Xb{{\bm X}}\def\Wb{{\bm W}}

\def\t{{\top}}

\def\rmspan{{\rm{span}}}

\def\tr{{\rm tr}}

\def\Xib{\mathbf{\Xi}}

\newcommand{\I}{{\cal I}}
\newcommand{\Ic}{{{\cal I}_{\rm nc}}}

\newtheorem{thm}{Theorem}[section]

\newtheorem{lem}[thm]{Lemma}

\begin{document}

\title{A robust RUV-testing procedure via $\gamma$-divergence}

\date{}
\author{Hung Hung\\[2ex]
\small Institute of Epidemiology and Preventive Medicine, National
Taiwan University, Taiwan}

\maketitle

\begin{abstract}
Identification of differentially expressed genes (DE-genes) is
commonly conducted in modern biomedical researches. However,
unwanted variation inevitably arises during
the data collection process, which could make the detection results
heavily biased. It is suggested to remove the unwanted variation
while keeping the biological variation to ensure a reliable analysis
result. Removing Unwanted Variation (RUV) is recently proposed for
this purpose by the virtue of negative control genes. On the other
hand, outliers are frequently appear in modern high-throughput
genetic data that can heavily affect the performances of RUV and its
downstream analysis. In this work, we propose a robust RUV-testing procedure via
$\gamma$-divergence. The
advantages of our method are twofold: (1) it does not involve any
modeling for the outlier distribution, which is applicable to
various situations; (2) it is easy to implement in the sense that
its robustness is controlled by a single tuning parameter $\gamma$
of $\gamma$-divergence, and a data-driven criterion is developed to
select $\gamma$. In the Gender Study, our method can successfully
remove unwanted variation, and is able to identify more DE-genes
than conventional methods.\\

\end{abstract}

\newpage

\section{Introduction}

Identification of differentially expressed genes (DE-genes) is
commonly conducted in modern biomedical researches. However,
unwanted variation (e.g., batch effects) inevitably arises during
the data collection process, which could make the detection results
heavily biased. Let $\{Y_{ij}\}_{j=1}^p$ be the measurements of $p$ genes
and $X_i$ be the covariate of interest (e.g.,
disease status) for the $i$-th subject,
$i=1,\ldots,n$. The problem of identifying DE-genes in the presence
of unwanted variation can be formulated as the following model:
\begin{eqnarray}
\Yb_j = \1_n\delta_j+\Xb\beta_j + \Wb\alphab_j+\varepsilonb_j,\quad
j=1,\ldots,p,\label{model}
\end{eqnarray}
where $\Yb_j=(Y_{1j},\ldots,Y_{nj})^\t$,
$\Xb=(X_1,\ldots,X_n)^\top$, $\beta_j$ is the effect size of the
$j$-th gene, $\Wb \in \mathbb{R}^{n\times k}$ represents possible
confounding factors (i.e., the sources of unwanted variation) with
the coefficient $\alphab_j\in \mathbb{R}^{k}$, $\delta_j$
is the intercept term, and $\varepsilonb_j=(\varepsilon_{1j},\ldots,\varepsilon_{nj})^\top\sim N(0,\sigma_j^2
\Ib_n)$ is the error term with the gene-specific variance $\sigma_j^2$. The dimension $k$ of $\Wb$ is the number of sources of
unwanted variation. For example, the batch effect with $(k+1)$
batches corresponds to $\Wb$ with each element being zero or one,
and the sum of each row of $\Wb$ is smaller than 1. In the rest of
discussion, we assume that $k$ is known, and discuss its selection
separately. We also assume $(\Xb,\Wb)$ have been centered such that
$\delta_j=E(Y_{ij})$. Model~(\ref{model}) can be expressed in
matrix form as
\begin{eqnarray}
\Yb = \1_n\deltab + \Xb\betab + \Wb\alphab +\varepsilonb
\quad \Longleftrightarrow \quad\Yb^* = \Xb\betab + \Wb\alphab
+\varepsilonb
 ,\label{model.m}
\end{eqnarray}
where $\Yb=[\Yb_1,\ldots,\Yb_p]_{n\times p}$,
$\deltab=[\delta_1,\ldots,\delta_p]_{1\times p}$,
$\betab=[\beta_1,\ldots,\beta_p]_{1\times p}$,
$\alphab=[\alphab_1,\ldots,\alphab_p]_{k\times p}$,
$\varepsilonb=[\varepsilonb_1,\ldots,\varepsilonb_p]_{n\times p}$, and
$\Yb^*=\Yb-\1_n\deltab$ is the (column) centered version of
$\Yb$. In practical implementation, $\Yb^*$ is replaced by
$\widehat\Yb^*=(\Ib_n-\frac{1}{n}\1_n\1_n^\top)\Yb$. We also use the
notation $\Yb_{\I}$ with the subindex of a set $\I$ to denote the
submatrix $[\Yb_j:j\in \I]$, which is taken from the columns of $\Yb$
in $\I$. A similar definition applies to $\Yb^*_{\I}$, $\betab_{\I}$, $\alphab_{\I}$, and $\varepsilonb_{\I}$.

Provided that $(\Yb,\Xb,\Wb)$ is observable, identification of
DE-genes under model~(\ref{model}) can be conducted by conventional
multiple hypothesis testing procedure using the p-values for testing
$H_0:\beta_j=0$. The problem is not straightforward since one usually can only observe $(\Yb,\Xb)$ while $\Wb$ is not
available. In the absence of $\Wb$, directly using p-value for
testing $H_0:\beta_j=0$ from fitting the model $\Yb_j =\1_n\delta_j
+\Xb\beta_j + \varepsilonb_j$ can produce biased result. To overcome
this problem, a two-stage procedure is commonly suggested:
\begin{enumerate}
\item[(S1)]
Using certain procedure to estimate $\Wb$ by $\widehat \Wb$.

\item[(S2)]
Estimate $\beta_j$ by fitting the model $\Yb_j =
\1_n\delta_j+\Xb\beta_j + \widehat\Wb\alphab_j+\varepsilonb_j$, $j=1,\ldots,p$.
\end{enumerate}
There are many methods developed to construct $\widehat\Wb$, including Price \emph{et al.} (2006), Leek and Storey (2007), Friguet, Kloareg, and Causeur (2009), Gagnon-Bartsch and
Speed (2012), Sun, Zhang, and Owen (2012), Gagnon-Bartsch, Jacob, and Speed (2013), Leek (2014), Risso \emph{et al.} (2014) among others. By considering $\widehat\Wb$, the
resulting p-value for $H_0:\beta_j=0$ in (S2) is expected to be
more reliable to identify DE-genes. Note that using an estimate of $\rmspan(\Wb)$ in (S2) suffices to
estimate $\beta_j$ consistently.

Among the existing methods, one branch of methods called \emph{Removing
Unwanted Variation} (RUV) (Gagnon-Bartsch and Speed, 2012; Gagnon-Bartsch, Jacob, and Speed, 2013) attracts much attentions recently, which estimates $\Wb$ by utilizing the
virtue of \emph{negative control genes}. Negative control
genes are genes that are known to have zero effects, i.e.,
$\beta_j=0$. Examples of negative control genes include house-keeping genes and spike-in controls (Lippa \emph{et al.}, 2010). As a result, variation that appears in negative control genes must be unwanted variation, and this fact provides a basis to estimate $\Wb$ without imposing further assumptions on the data. There are many versions of RUV, and we review the commonly used RUV2 and RUV4 below. Let $\Ic$ be the index set of negative control genes. It follows from (\ref{model.m}) that
\begin{eqnarray}
\Yb^*_{\Ic} = \Wb\alphab_\Ic+\varepsilonb_\Ic. \label{model.c}
\end{eqnarray}
Based on (\ref{model.c}), the RUV2 starts from applying singular
value decomposition (SVD) to obtain $\Ub_k$, which consists of the
leading $k$ left singular vectors of $\widehat\Yb_{\Ic}^*$. The RUV2
estimator of $\Wb$ is then given by $\widehat\Wb_{\rm ruv2}=\Ub_k$.
The RUV4 uses negative control genes with a rotation of $\Yb^*$. Let
$[\Rb_0,\Rb_1]\in \mathbb{R}^{n\times n}$ be the orthogonal matrix
such that $\rmspan(\Rb_1)=\rmspan(\Xb)$. It implies from
(\ref{model.m}) that
\begin{eqnarray}
\Rb_0^\top \Yb^* &=& (\Rb_0^\top\Wb)\alphab+\Rb_0^\top\varepsilonb\label{ruv4.1}\\
\Rb_1^\top \Yb_{\Ic}^* &=&
(\Rb_1^\top\Wb)\alphab_{\Ic}+\Rb_1^\top\varepsilonb_{\Ic} \label{ruv4.2}.
\end{eqnarray}
Applying SVD on $\Rb_0^\top \widehat\Yb^*$ to obtain $(\Ub_k,\Vb_k)$
such that $\Ub_k^\top \Ub_k=\Ib_k$ and $\Ub_k\Vb_k^\top$ is the best
rank-$k$ approximation of $\Rb_0^\top \widehat\Yb^*$. The RUV4
starts from (\ref{ruv4.1}) to estimate $(\Rb_0^\top\Wb, \alphab)$ by
$\widehat{\Rb_0^\top\Wb}=\Ub_k$ and $\widehat\alphab=\Vb_k^\t$. Second,
from (\ref{ruv4.2}) with $\alphab_{\Ic}$ being replaced by
$\widehat{\alphab}_{\Ic}$, an estimate of $\Rb_1^\top\Wb$ is obtained
as $\widehat{\Rb_1^\top\Wb}=(\Rb_1^\top
\widehat\Yb_{\Ic}^*)\widehat{\alphab}_{\Ic}^\top(\widehat{\alphab}_{\Ic}
\widehat{\alphab}_{\Ic}^\top)^{-1}$ from LSE. Finally, the RUV4
estimator of $\Wb$ is constructed to be $\widehat\Wb_{\rm
ruv4}=\Rb_0\cdot\widehat{\Rb_0^\top\Wb}+\Rb_1\cdot\widehat{\Rb_1^\top\Wb}$.
See also Gerard and Stephens (2017) for an insightful review of RUV
methods.

A key step of the above RUV procedures is SVD, which is sensitive to
outliers (i.e., when $\Yb$ is contaminated) and the resulting
estimation of $\Wb$ in (S1) can be heavily biased. The contaminated
$\Yb$ can also affect the estimation of $\beta_j$ to produce
unreliable p-values in (S2). Outliers, however, are frequently
encountered in modern high-throughput genetic studies. Therefore, both
the robustness for the RUV stage (S1) and the testing stage (S2) are
critical to ensure a reliable analysis result. The aim of this study
is to propose a robust RUV-testing procedure of (S1)-(S2), via
utilizing the robust $\gamma$-divergence. An important merit of our
method is that the robustness comes from the robust minimum
$\gamma$-divergence estimation criterion, which does not involve any
modeling for the outlier distribution (see
Section~\ref{sec.gamma_div} for details). That is, our method is
flexible to adapt to various situations. We will also
see that the robustness of our proposal is achieved by the concept
of ``weighted by the model'', where the weight is induced from
model~(\ref{model}) directly. As a result, our method is easy
to implement in the sense that, the robustness is merely controlled
by a single tuning parameter $\gamma$ of $\gamma$-divergence, and a
data-driven selection criterion of $\gamma$ is developed.

\section{Method}

\subsection{A review of $\gamma$-divergence}\label{sec.gamma_div}

For any $\gamma>0$, the $\gamma$-divergence between the model
distribution $f_\theta$ (indexed by the parameter $\theta$) and the
true distribution $g$ is defined to be
\begin{eqnarray}
\mathcal{D}_{\gamma}(g,f_\theta)=\frac{1}{\gamma(\gamma+1)}
\left\{\|g\|_{\gamma+1} -\int
\left(\frac{f_\theta}{\|f_\theta\|_{\gamma+1}}\right)^\gamma
g\right\},\label{gamma_div}
\end{eqnarray}
where $\|f_\theta\|_{\gamma+1}=(\int
f_\theta^{\gamma+1})^{\frac{1}{\gamma+1}}$. Note that
$\mathcal{D}_{\gamma}(g,f_\theta)=0$ if and only if
$f_\theta=\lambda g$ for some $\lambda>0$. In the limiting case,
$\lim_{\gamma\to 0}\mathcal{D}_\gamma$ reduces to the KL-divergence.
The minimum $\gamma$-divergence estimation criterion estimates
$\theta$ by
\begin{eqnarray}
\argmin_\theta \mathcal{D}_{\gamma}(g,f_\theta) =\argmax_\theta\int
\left(\frac{f_\theta}{\|f_\theta\|_{\gamma+1}}\right)^\gamma
g.\label{gamma.div.est}
\end{eqnarray}
When the model $f_\theta$ is correctly specified (i.e.,
$g=f_{\theta_0}$ for some $\theta_0$), one has
$\theta_0=\argmin_\theta \mathcal{D}_{\gamma}(g,f_\theta)$ which
supports the consistency of (\ref{gamma.div.est}). At the sample
level, $g$ is replaced by the empirical distribution of the data to
obtain the estimate of $\theta_0$.

In the presence of outliers, data is no longer generated from
$f_{\theta_0}$ but the mixture $cf_{\theta_0}+(1-c)h$, where $h$ is
the outlier distribution and $c$ is the mixing proportion. The
robustness of $\gamma$-divergence can be observed by noting that
\begin{eqnarray}
D_\gamma\{cf_{\theta_0}+(1-c)h,f_\theta\}= c\,
D_\gamma(f_{\theta_0},
f_\theta)-\frac{1-c}{\gamma(\gamma+1)}\int\left(\frac{f_{\theta}}{\|f_\theta\|_{\gamma+1}}\right)^\gamma
h+\tau, \label{div.gamma.contaminated}
\end{eqnarray}
where $\tau$ is a constant that is independent of $\theta$. Thus,
provided that
$\int(\frac{f_{\theta}}{\|f_\theta\|_{\gamma+1}})^\gamma h$ is small
at $\theta\approx\theta_0$, the minimizer of
(\ref{div.gamma.contaminated}) is approximately equivalent to the
minimizer of $D_\gamma(f_{\theta_0}, f_\theta)$, which does not
depend on $(c,h)$. See Fujisawa and Eguchi (2008) and Kanamori and
Fujisawa (2015) for more discussions about the robustness of
$\gamma$-divergence.

\subsection{The robust RUV procedure: $\gamma$-RUV}

The target of the RUV procedure (S1) is the matrix $\Wb$, which is
treated as the unknown parameter in this subsection. The basic idea
of our robust RUV method is to re-formulate model~(\ref{model.c}) as a location-scale model, from which we can apply $\gamma$-divergence to robustly
estimate $\Wb$. To develop our method, we assume the following two conditions on model~(\ref{model.c}) as our ``working model'' for negative
control genes:
\begin{enumerate}
\item[(C1)]
$\alphab_j$ follows an arbitrary distribution with mean
$\mub_\alphab\in \mathbb{R}^k$ and non-singular covariance matrix
$\Sigmab_\alphab \in \mathbb{R}^{k\times k}$, $j\in\Ic$.
\item[(C2)]
$\sigma_j^2=\sigma^2$, $j\in\Ic$.
\end{enumerate}
Assumption (C1) implies that model~(\ref{model.c}) can be
equivalently expressed as
\begin{eqnarray}
\Yb_j^* = \mub + \Gammab\nub_j+\varepsilonb_j,\quad j\in \Ic,
\label{model.pca}
\end{eqnarray}
where $\mub=\Wb\mub_\alphab$, $\Gammab$ is the leading $k$ eigenvectors of $\Wb\Sigmab_\alphab\Wb^\top$, and $\nub_j=\Gammab^\top\Wb(\alphab_j-\mub_\alphab)$
is a zero mean
random vector with diagonal covariance matrix.
Assumption (C2) further implies that $E(\Yb_j^*)=\mub$ and
$\cov(\Yb_j^*)=\Sigmab=\Gammab{\rm cov}(\nub_j)\Gammab^\top+\sigma^2\Ib_n$
with $\Gammab$ being the leading eigenvectors of $\Sigmab$.
Note that $\rmspan([\mub,\Gammab])=\rmspan(\Wb)$ by the
definitions of $(\mub,\Gammab)$ and, hence, a basis of
$\rmspan(\Wb)$ can be estimated by $[\mub,\Gammab]$. It also implies
that a robust RUV procedure can be established by any robust
estimator of $(\mub,\Sigmab)$ under model~(\ref{model.pca}). There exist many robust methods for
location and scale parameters in the literatures. We adopt the
minimum $\gamma$-divergence estimation criterion
(\ref{gamma.div.est}) to estimate $(\mub,\Sigmab)$ as described below. Let $f_\theta$ with
$\theta=(\mub,\Sigmab)$ be the pdf of the multivariate normal
distribution with mean $\mub$ and covariance matrix $\Sigmab$
induced from model~(\ref{model.pca}). Substituting
$f_\theta$ into (\ref{gamma.div.est}), replacing $g$ with the
empirical distribution of $\{\Yb_j^*\}_{j\in \Ic}$,  and taking differentiation with respect to $\theta$, we obtain the
estimating equations of $(\mub,\Sigmab)$ to be
\begin{eqnarray}
\mub&=&\frac{\sum_{j\in \Ic}f_{\theta}^\gamma(\Yb_j^*)\Yb_j^*}
{\sum_{j\in \Ic}f_{\theta}^\gamma(\Yb_j^*)}\label{gamma.mu}\\
\Sigmab&=&(\gamma+1)\frac{\sum_{j\in
\Ic}f_{\theta}^\gamma(\Yb_j^*)(\Yb_j^*-\mub)(\Yb_j^*-\mub)^\top}{\sum_{j\in
\Ic}f_{\theta}^\gamma(\Yb_j^*)}.\label{gamma.sigma}
\end{eqnarray}
The solution $\widehat\theta=(\widehat\mub, \widehat\Sigmab)$ of (\ref{gamma.mu})-(\ref{gamma.sigma})
provides a robust estimate of $(\mub,\Sigmab)$.

With the above derivations, the robust estimator of $\Wb$ is proposed to be
\begin{eqnarray}
\widehat\Wb_\gamma=[\widehat\mub,\widehat\Gammab_k],\label{W.gamma}
\end{eqnarray}
where $\widehat\Gammab_k$ is the leading $k$ eigenvectors of
$\widehat\Sigmab$, and $(\widehat\mub, \widehat\Sigmab)$ are the
solutions of (\ref{gamma.mu})-(\ref{gamma.sigma}) with $\Yb_j^*$
being replaced by $\widehat \Yb_j^*$. We call the RUV
procedure using $\widehat\Wb_\gamma$ in (S1) as $\gamma$-RUV.



\subsection{The robust testing procedure: $\gamma$-LSE}\label{sec.2.test}

Although $\gamma$-RUV can robustly estimate $\Wb$ in (S1), outliers
can still affect the performance of the
downstream analysis in (S2). To overcome this problem, we use
$\gamma$-divergence to develop a robust testing procedure for
$H_0:\beta_j=0$ under model~(\ref{model}). In this section, $\Wb$ is
assumed to be known and $\alphab_j$ is treated as the unknown but
fixed parameter. In practical
implementation, $\Wb$ will be replaced by
$\widehat\Wb_\gamma$ when combined with $\gamma$-RUV.

For any $1\le j\le p$, we start by re-expressing model~(\ref{model}) as
\begin{eqnarray}
Y_{ij}=\etab_j^\top \Zb_i+\varepsilon_{ij},\quad
i=1,\ldots,n,\label{model.testing}
\end{eqnarray}
where $\etab_j=(\delta_j,\beta_j,\alphab_j^\top)^\top$ and
$\Zb_i=(1,X_i,\Wb_i^\top)^\top$ with $\Wb_i^\top$ being the $i$-th row of
$\Wb$. Model~(\ref{model.testing}) is equivalent to assuming $Y_{ij}|\Zb_i\sim f_{\theta_j}(y|\Zb_i)$, where
$f_{\theta_j}(y|\zb)$ with $\theta_j=(\etab_j,\sigma_j^2)$ is
the pdf of the normal distribution with mean $\etab_j^\top \zb$ and variance $\sigma_j^2$.
Based on the data $\{(Y_{ij},\Zb_i)\}_{i=1}^n$, we follow Hung, Zou, and Huang (2017) to estimate $\theta_j$ via the modified minimum
$\gamma$-divergence estimation criterion (to adapt to a regression
setting):
\begin{eqnarray}
\max_{\theta_j}
\frac{1}{n}\sum_{i=1}^n\left(\frac{f_{\theta_j}(Y_{ij}|\Zb_i)}{\|f_{\theta_j}(\cdot|\Zb_i)\|_{\gamma+1}}\right)^\gamma.
\end{eqnarray}
Direct differentiation gives the estimating equations of $\theta_j$
to be
\begin{eqnarray}
\etab_j &=& (\Zb^\top \Omegab_{\theta_j}^\gamma\Zb)^{-1}\Zb^\top \Omegab_{\theta_j}^\gamma\Yb_j\label{gamma.lse.beta}\\
\sigma_j^2 &=&
(\gamma+1)\frac{(\Yb_j-\Zb\etab_j)^\top\Omegab_{\theta_j}^\gamma(\Yb_j-\Zb\etab_j)}{\tr(\Omegab_{\theta_j}^\gamma)}\label{gamma.lse.sigma},
\end{eqnarray}
where $\Omegab_{\theta_j}={\rm
diag}\{f_{\theta_j}(Y_{1j}|\Zb_1),\ldots,f_{\theta_j}(Y_{nj}|\Zb_n)\}$
and $\Zb=[\Zb_1,\ldots,\Zb_n]^\top$. Denote the solution by
$\widehat\theta_{j}=(\widehat\etab_j,\widehat\sigma_{j}^2)$ with
$\widehat\etab_j=(\widehat\delta_{j},\widehat\beta_{j},\widehat\alphab_{j}^\top)^\top$,
which we call $\gamma$-LSE of $\theta_j$.

The asymptotic covariance matrix can be empirically estimated by
$\widehat\Sb_{j}$. The null distribution of $\widehat\beta_j$ can
then be approximated by $N(0,[\widehat\Sb_{j}]_{\beta_j}/n)$, where
$[\widehat\Sb_{j}]_{\beta_j}$ is the diagonal element of
$\widehat\Sb_{j}$ that corresponds to $\beta_{j}$. Consequently, a
robust p-value for testing $H_0:\beta_j=0$ is constructed to be
\begin{eqnarray}\label{gamma.pj}
\rho_{j} =1-
F_{\chi^2_1}\left(n\widehat\beta_{j}^2/[\widehat\Sb_{j}]_{\beta_j}\right),
\end{eqnarray}
where $F_{\chi^2_1}$ is the cdf of the chi-square distribution with
degrees of freedom 1.

Based on the developed $\gamma$-RUV and $\gamma$-LSE, the robust identified index set of DE-genes
is proposed to be
\begin{eqnarray}
\widehat\I_{\rm de}=\left\{j:\rho_{j}<\frac{\alpha}{p}\right\}\label{gamma.Ide}
\end{eqnarray}
with the family-wise error rate being controlled at $\alpha$ by
Bonferroni correction, where $\rho_j$ is obtained from (\ref{gamma.pj}) with $\Wb$ being replaced by $\widehat\Wb_\gamma$. We remind the readers that (\ref{gamma.Ide}) can be used with any estimation method of $\Wb$ and is not limited to $\gamma$-RUV. However, a non-robust estimate of $\Wb$ will make the p-values unreliable, even combined with the robust $\gamma$-LSE.
The robustness of both RUV and testing stages should be taken into account in order to ensure a reliable analysis result.

\section{Simulation Studies}

\subsection{Simulation settings}

We generate each element of $\Xb$ from Bernoulli($0.5$). The latent
source of unwanted variation $\Wb=[\Wb_{(1)}, \Wb_{(2)}]$ consists
of two parts, where $\Wb_{(1)}\in \mathbb{R}^{n\times 4}$
represents the batch effect with $5$ batches, and $\Wb_{(2)}\in
\mathbb{R}^{n\times 3}$ represents the random
unwanted variation. We generate each row of $\Wb_{(1)}$ from
Multinomial($1, \frac{1}{5}\1_{5}$), and set $\Wb_{(2)}=2\Xb\zetab+\Eb$, where $\zetab\in
\mathbb{R}^{1\times 3}$ follows the uniform distribution on the $3$-dimensional sphere, and
each element of $\Eb\in
\mathbb{R}^{n\times 3}$ follows $N(0,1)$. It
gives the proportion of variation of $\Wb$ explained by $\Xb$ to be
$25\%$. Given $(\Xb, \Wb)$, the
uncontaminated data $\Yb_0$ is generated from model~(\ref{model.m})
as $\Yb_0=\1_n\deltab+\Xb\betab+\Wb\alphab+\varepsilonb$,
where each element of $\deltab$ is generated from $N(0,2^2)$. For
$\betab$, we set the first $100$ genes to be DE-genes with effect
sizes $\beta_j\sim N(1,0.2^2)$ and set $\beta_j=0$ for $j>100$. For
$\alphab$, each $\alphab_j$ is generated from $N(\0,\Ib_7)$,
$j=1,\ldots,p$. For the error term $\varepsilonb$, each
$\varepsilonb_j$ is generated from $N(\0,\sigma_j^2\Ib_n)$, where
$\sigma_j^2$ follows the inverse-gamma distribution with
parameters $(3,0.5)$ such that $E(\sigma_j^2)={\rm
var}(\sigma_j^2)=1$, $j=1,\ldots,p$.


To mimic the presence of outliers, we generate the outlier matrix by
\begin{eqnarray}
\Ob=[\Xb,\Wb_{(1)}]\zetab_{\rm o}+\Eb_{\rm o},\label{ol}
\end{eqnarray}
where each element of $\zetab_{\rm o}\in \mathbb{R}^{5\times p}$
is generated from $N(0,\sigma_{\rm o}^2)$, and each element of
$\Eb_{\rm o}\in \mathbb{R}^{n\times p}$ is generated from $N(0,1)$.
We then randomly select $p(1-\sqrt{\pi_{\rm o}})$ columns of $\Ob$
and set their values to be zeros. For the rest non-zero
$p\sqrt{\pi_{\rm o}}$ columns, each element is further set to be
zero with probability $(1-\sqrt{\pi_{\rm o}})$. Finally, the
observed (contaminated) data is $\Yb=\Yb_0+\Ob$ with about $100\pi_{\rm o}\%$ elements being contaminated.

For each simulation run, we use the last 200 genes of $\Yb$ as the
negative control genes, i.e., $\Ic=\{p-199,p-198,\ldots,p\}$, to
obtain $\widehat\Wb_\gamma$ by $\gamma$-RUV, and then obtain the
robust p-values $\rho_j$'s by fitting $\gamma$-LSE on
$(\Yb,\Xb,\widehat\Wb_\gamma)$. We also report the results from
fitting LSE on $(\Yb,\Xb,\widehat\Wb_\gamma)$ to demonstrate the
influence of outliers in the stage of testing. Since a good
RUV-testing procedure is expected to produce p-values that can well
separate DE-genes and non DE-genes, the AUC value from the two
groups $\{\rho_j:1\le j\le 100\}$ and $\{\rho_j:101\le j\le p-200\}$
is used as the performance measure (exclude negative control genes
from calculating AUC). A large AUC value indicates a good
performance of the RUV-testing procedure. Besides AUC, the true
positive (TP) and false positive (FP) values from the identified set
of DE-genes $\widehat\I_{\rm de}$ under $\alpha=0.05$ are reported.
The simulation results from RUV2 and RUV4 are also reported for
comparisons. Moreover, we report the results without the RUV stage
(i.e., ignore the effect of $\Wb$, denoted as ``Ignore-$\Wb$''), and
the results from $(\Yb_0,\Xb,\Wb)$ (i.e., use the uncontaminated
data $\Yb_0$ and true $\Wb$, denoted as ``Ideal'') as the wort and
best situations, respectively. It gives a total of 10 RUV-testing
procedures (5 RUV strategies combined with 2 testing methods). We
use the term ``A[+B]'' to denote a combination of RUV-testing
procedure, where ``A'' is the RUV strategy used in (S1) and ``B'' is
the testing method used in (S2). Simulation results are reported
with 100 replicates.

\subsection{Simulation results}

\begin{figure}[!ht]
\centering
\includegraphics[width=5in]{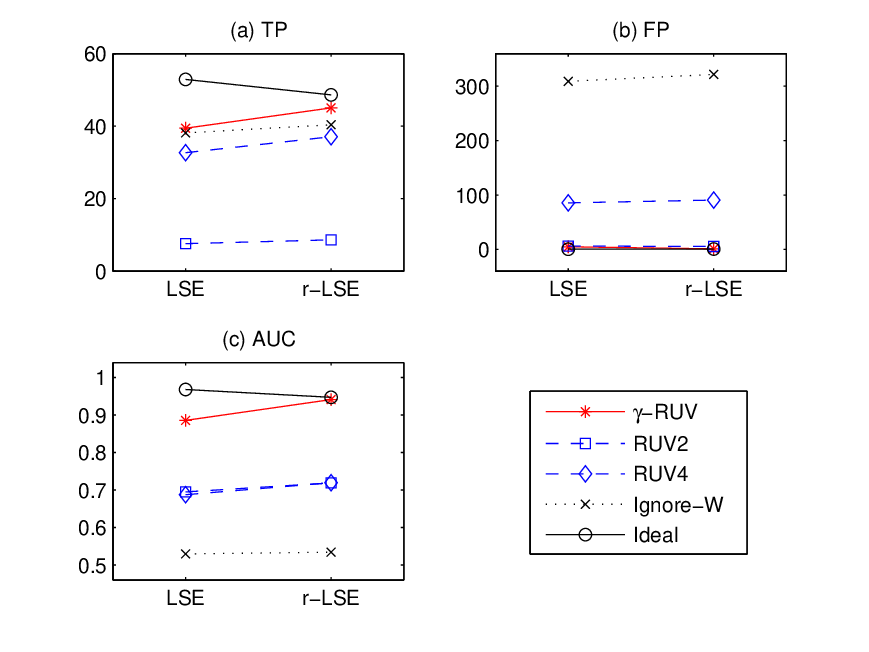}
\caption{The simulation results of $\gamma$-RUV, RUV2, and RUV4
combined with $\gamma$-LSE and LSE. Each line represents an RUV method combined with LSE (the left point) and $\gamma$-LSE (the right point). (a) The mean of TP. (b) The
mean of FP. (c) The mean of AUC.} \label{fig.sim1}
\end{figure}

Simulation results with $k=8$ under
$(n,p)=(100,1000)$ and $(\pi_{\rm
o},\sigma_{\rm o})=(0.05,20)$ are placed in Figure~\ref{fig.sim1}. We
first evaluate the performances of the RUV methods by inspecting the
simulation results under the conventional LSE fitting. Observe the
large differences between (TP, FP, AUC) of Ignore-$\Wb$[+LSE] and
Ideal[+LSE], it indicates the potential of RUV to improve the detection abilities. However, both RUV2[+LSE] and RUV4[+LSE] can only produce AUC values around 0.7, and there still has a large space for improvement in comparison with Ideal[+LSE]. On the other hand,
$\gamma$-RUV[+LSE] can achieve AUC value about 0.9, which is
significantly larger than RUV2[+LSE] and
RUV4[+LSE]. We also detect larger (lower) values of TP (FP) for
$\gamma$-RUV[+LSE]. It indicates that the conventional RUV procedures can be sensitive to the presence of outliers, while the proposed $\gamma$-RUV is able to
correctly recover $\Wb$ without being affected much by outliers.

We next evaluate the performance of $\gamma$-LSE. It can be seen
that $\gamma$-RUV[+$\gamma$-LSE] has a larger AUC value than
$\gamma$-RUV[+LSE]. It demonstrates that $\gamma$-RUV can only
produce a robust estimate of $\Wb$, while outliers can still affect
the detection powers in (S2). As a result, we can detect a further
improvement from the robust $\gamma$-LSE fitting over LSE. It suggests that a combination of $\gamma$-RUV and
$\gamma$-LSE, which can robustly estimate both $\Wb$ and $\beta_j$,
is preferred in practice. One can observe that all the lines
connecting the results of LSE and $\gamma$-LSE in Figure~\ref{fig.sim1} (a) and (c) show increasing trends
(i.e., an improvement from fitting $\gamma$-LSE over LSE), except for the
case of Ideal. Note that LSE is the most
efficient method in estimating $\beta_j$ when the model is correctly specified, and it is reasonable that
Ideal[+LSE] outperforms Ideal[+$\gamma$-LSE].

\begin{figure}[!ht]
\centering
\includegraphics[width=5in]{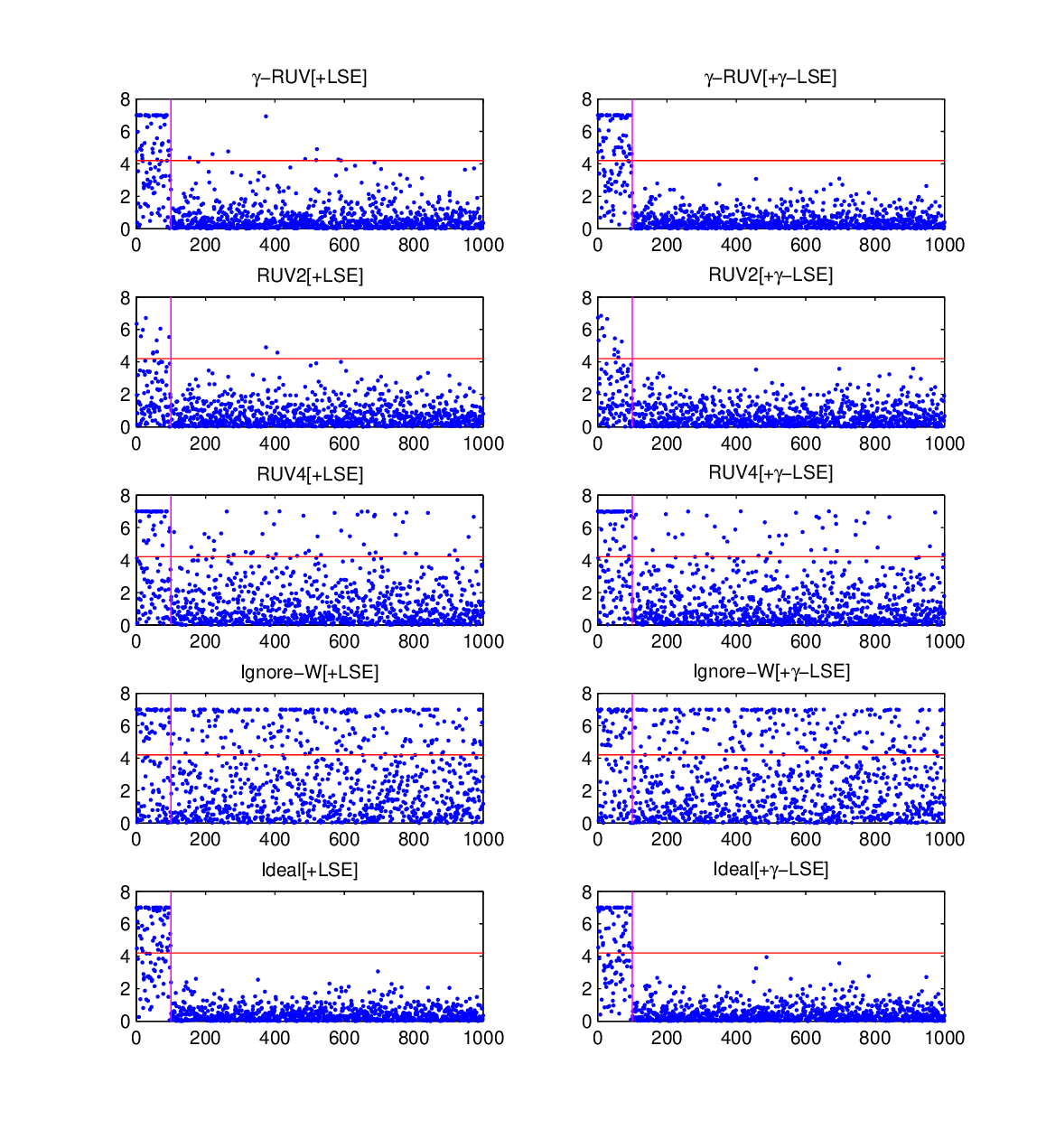}
\caption{One realization of the $-(\log_{10})$-transformed p-values for testing
$H_0:\beta_j=0$, $j=1,\ldots,1000$, from different RUV-testing procedures. The first 100
genes (in the left of the vertical line) are DE-genes, and the rest are non-DE genes. The
horizontal line represents the critical value under the family-wise
error rate 0.05.} \label{fig.sim2}
\end{figure}

To have more insights about the advantages of $\gamma$-RUV and
$\gamma$-LSE, we report the $(-\log_{10})$-transformed p-values from one simulation run in Figure~\ref{fig.sim2}. It can be
seen that the p-values from RUV2[+LSE], RUV4[+LSE], and
Ignore-$\Wb$[+LSE] cannot well separate DE-genes and non DE-genes,
while a more clear separation is observed for $\gamma$-RUV[+LSE],
indicating the robustness of $\gamma$-RUV.
One can also observe that $\gamma$-RUV[+$\gamma$-LSE] tends to produce larger (smaller) $(-\log_{10})$-transformed
p-values for DE-genes (non-DE genes) than $\gamma$-RUV[+LSE], indicating the robustness of $\gamma$-LSE. More specifically,
there are 9 genes that are falsely identified by
$\gamma$-RUV[+LSE], but the problem vanishes for
$\gamma$-RUV[+$\gamma$-LSE]. It
demonstrates the superiority of $\gamma$-RUV[+$\gamma$-LSE] in
the presence of outliers.

\section{Gender Study}

The Gender Study (GS) data is originally from Vawter \emph{et
al.} (2004) to investigate the effects of genes on brain under
different gender statuses. The data collects for each of 10
patients (5 male and 5 female) gene expressions from three regions
of brain (the anterior cingulate cortex, the dorsolateral prefrontal
cortex, and the cerebellar hemisphere), where each chip is carried
out in one of the three labs (UC Irvine, UC Davis, and University of
Michigan, Ann Arbor) and one of the two platforms (Affymetrix
HG-U95A and Affymetrix HG-U95Av2). The original data can be found in
Gene Expression Omnibus (GSE2164). In our analysis, we use the same
data with Gagnon-Bartsch and Speed (2012) after
BG-QN-LS preprocessing, which consists of 12600 gene expressions for
each of 84 chips as $\Yb$. The factor of interest $X$ is gender, and
the 488 genes on the X/Y chromosome are treated as DE-genes. The
negative control genes are 799 house-keeping (HK) genes from Eisenberg, and
Levanon (2003). Gagnon-Bartsch and Speed (2012) have analyzed the GS
data to evaluate the performance of RUV2, by inspecting the number
of the identified DE-genes. Since gene expression data can subject
to the problem of outliers, the aim of
our analysis is to re-analyze the GS data by
$\gamma$-RUV and $\gamma$-LSE (with $\ell=0.95$).

\begin{figure}[!ht]
\centering
\includegraphics[width=3in]{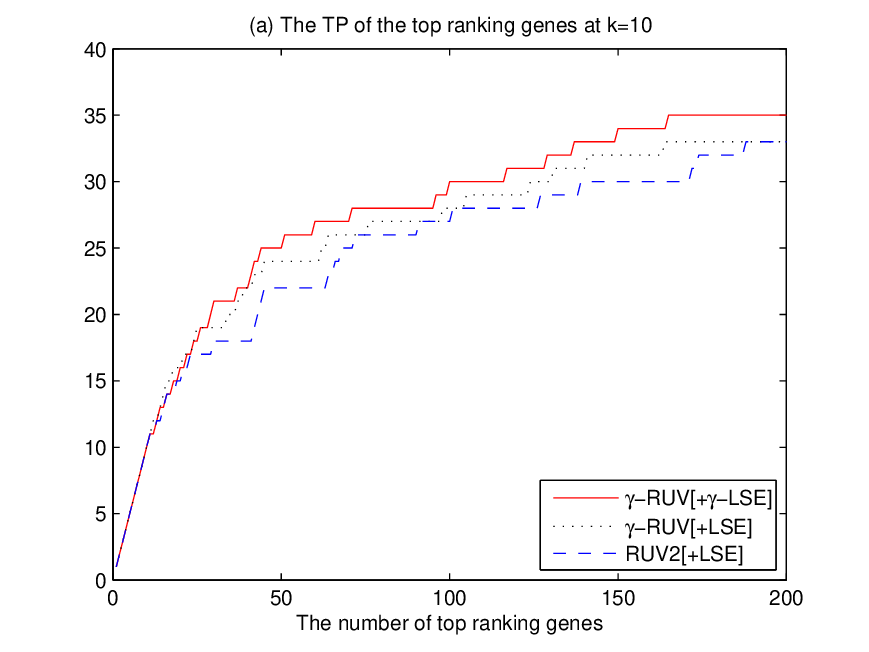}
\includegraphics[width=3in]{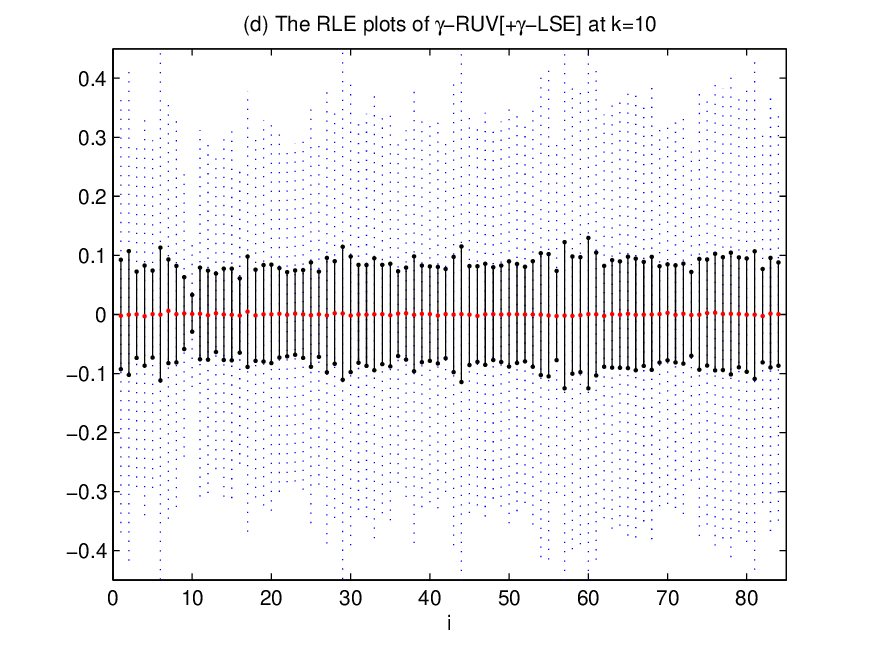}
\includegraphics[width=3in]{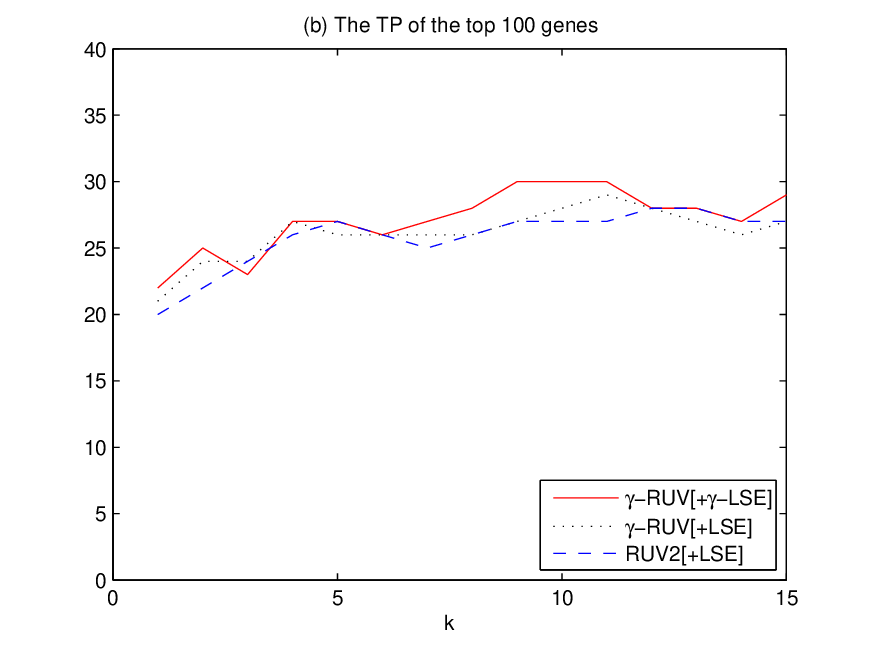}
\includegraphics[width=3in]{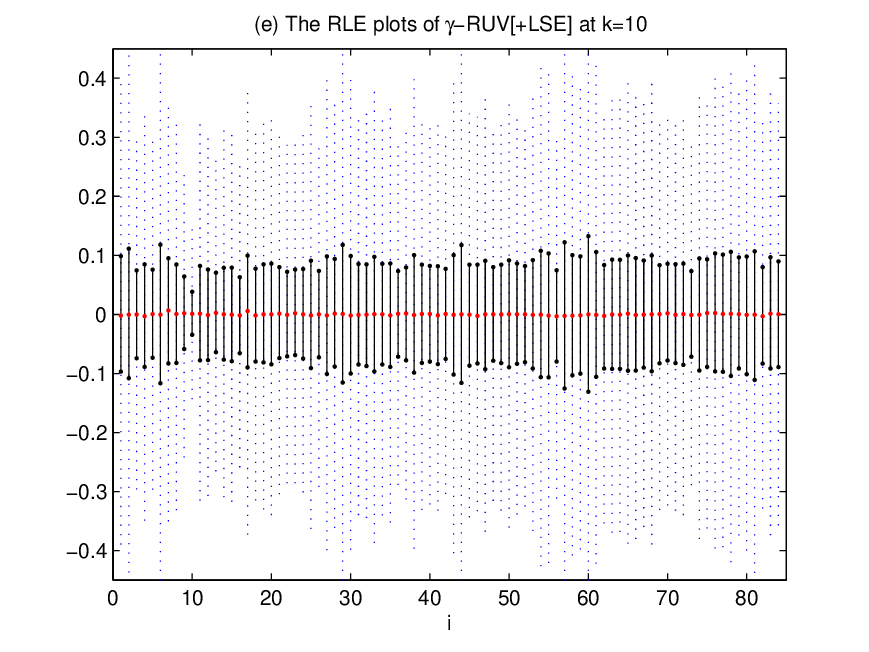}
\includegraphics[width=3in]{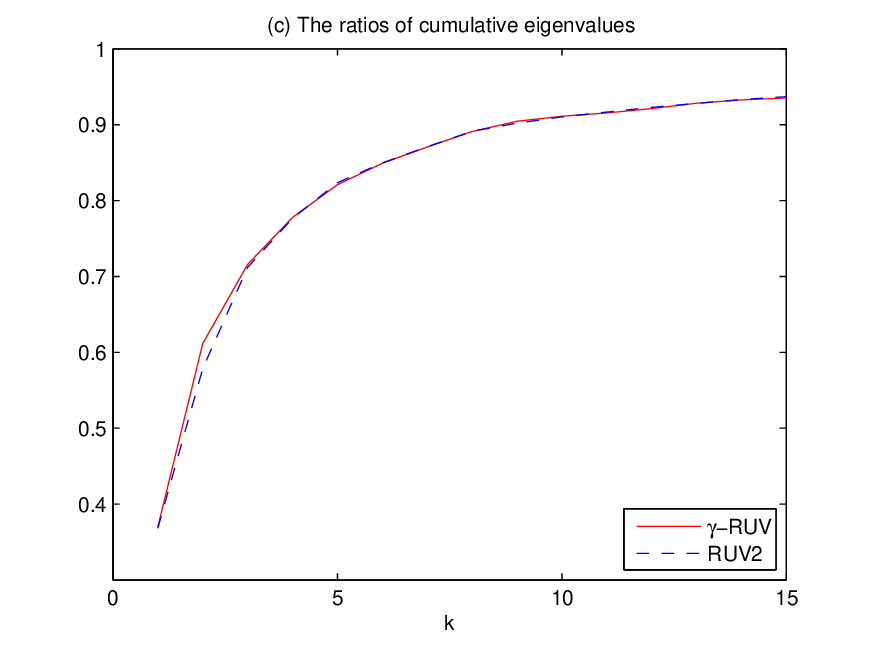}
\includegraphics[width=3in]{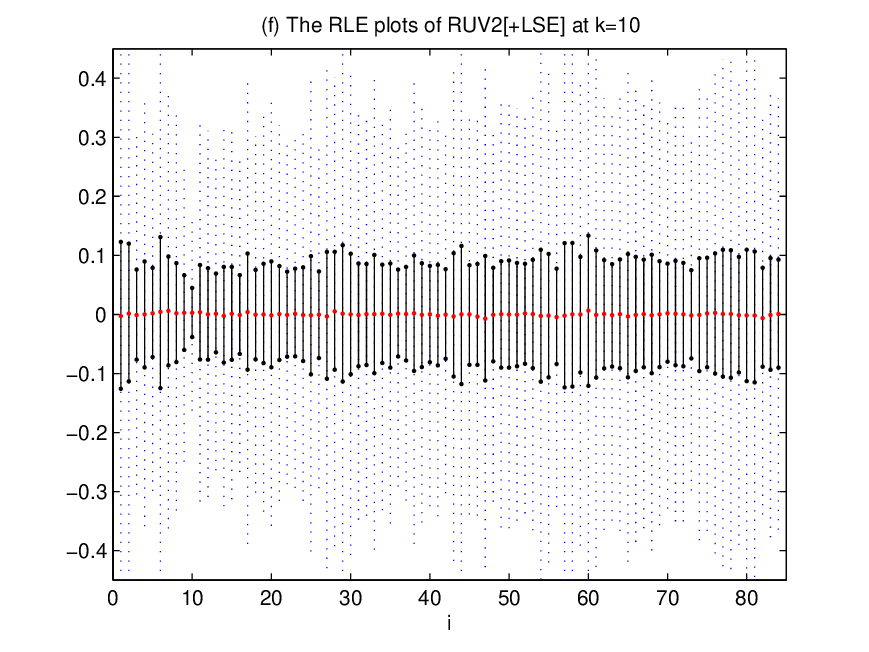}
\caption{The analysis results of the GS data. (a) The TP of the top ranking genes at $k=10$. (b) The TP of the top 100 genes at different $k$. (c) The ratios of cumulative eigenvalues at different $k$. (d) The RLE plots of $\gamma$-RUV[+$\gamma$-LSE] at $k=10$. (e) The RLE plots of $\gamma$-RUV[+LSE] at $k=10$. (f) The RLE plots of RUV2[+LSE] at $k=10$.} \label{fig.GS}
\end{figure}

Figure~\ref{fig.GS}~(a) reports the TP of the top ranking genes
at $k=10$. One can see that the best performer is
$\gamma$-RUV[+$\gamma$-LSE], followed by $\gamma$-RUV[+LSE] and
RUV2[+LSE]. It shows that both $\gamma$-RUV and $\gamma$-LSE have their own
contribution to the identification of DE-genes. Moreover, the differences between the $\gamma$-divergence based
methods and the conventional method implies the existence of some
abnormal data points, which can be influential to the analysis
result if not properly handled.

The performance of a RUV-testing procedure can be evaluated via the
relative log expression (RLE) plots. For the $i$-th chip with
$\{Y_{ij}\}_{j=1}^p$, its RLE plot is the boxplot of $\{
Y_{ij}-m_j\}_{j=1}^p$, where $m_j$ is the median of
$\{Y_{ij}\}_{i=1}^n$. A chip with good quality (i.e., when the
unwanted variation is properly removed) generally has the RLE plot
centered around zero with the interquartile range (IQR) being
smaller than 0.2. See Bolstad \emph{et al.} (2005) and Gandolfo and Speed (2017) for more
descriptions about the RLE plots. Figure~\ref{fig.GS}~(d)-(f) give
the RLE plots of the adjusted data
$(\Yb-\widehat\Wb\widehat\alphab)$, where
$(\widehat\Wb,\widehat\alphab)$ is obtained from a certain
RUV-testing procedure with $k=10$, and the corresponding mean IQR
over 84 chips are reported in Table~\ref{tab.lre}. One can see that
all methods produce satisfactory RLE plots, which indicates that the
outlyingness is not severe for the GS data. However,
$\gamma$-RUV[+$\gamma$-LSE] is still able to improve the
conventional RUV-testing procedure (see Figure~\ref{fig.GS}~(a)), and achieves the smallest mean
IQR value $0.172$.

Figure~\ref{fig.GS}~(b) reports the TP of the top 100 genes at
different $k$ values, and the corresponding ratios of cumulative
eigenvalues are plotted in Figure~\ref{fig.GS}~(c). The ratios of
cumulative eigenvalues support the selection of $k=10$ such that the proportion
of explained variation is about $90\%$. Note that in Figure~\ref{fig.GS} (b), the TP value of
$\gamma$-RUV[+$\gamma$-LSE] achieves the maximum value 30 at $k=10$, and has similar values for
$k$ near 10. It indicates the appropriateness for the selection of
$k$ by inspecting the ratios of cumulative eigenvalues of the robust $\widehat\Sigmab$.

\begin{figure}[!ht]
\centering
\includegraphics[width=3in]{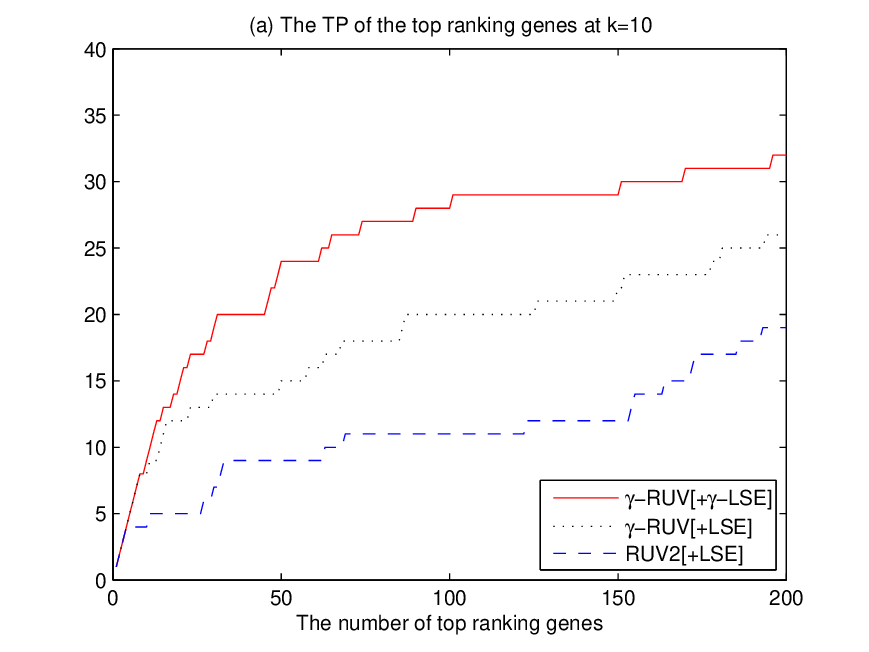}
\includegraphics[width=3in]{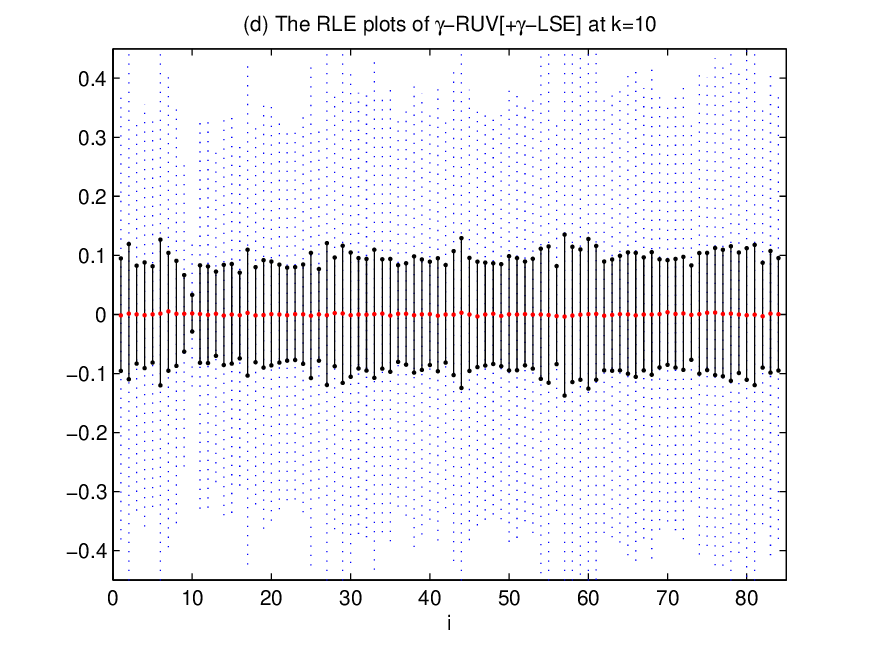}
\includegraphics[width=3in]{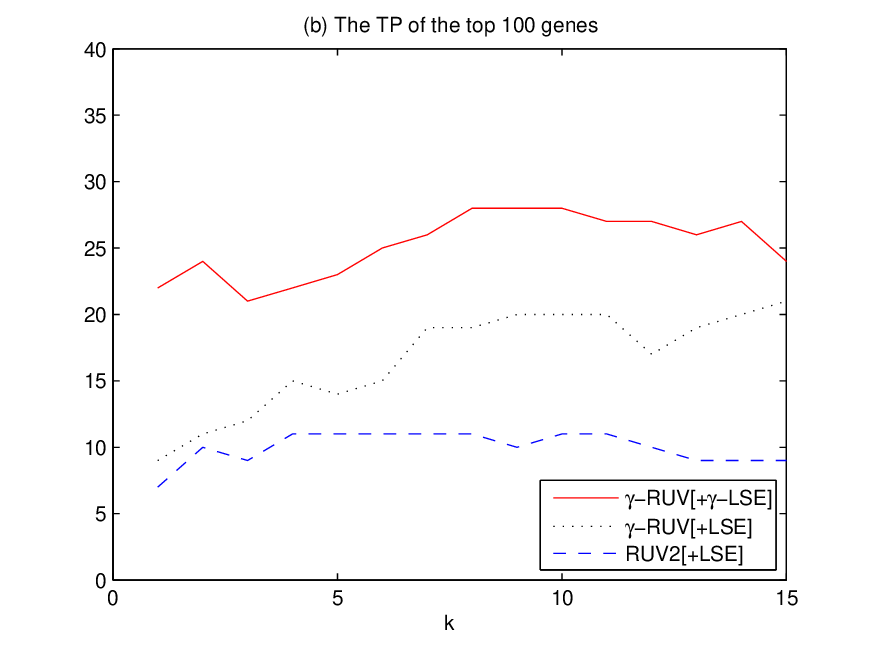}
\includegraphics[width=3in]{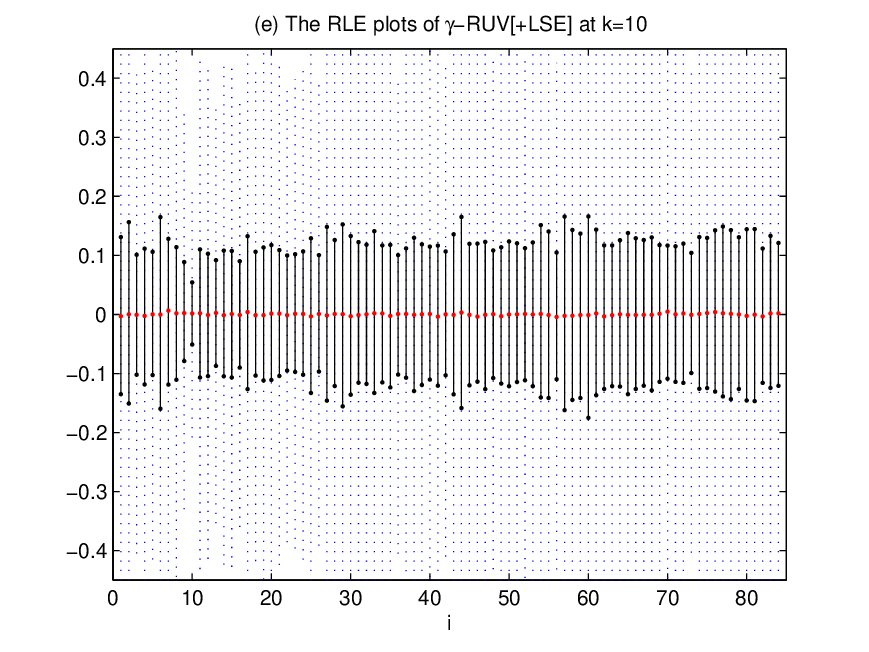}
\includegraphics[width=3in]{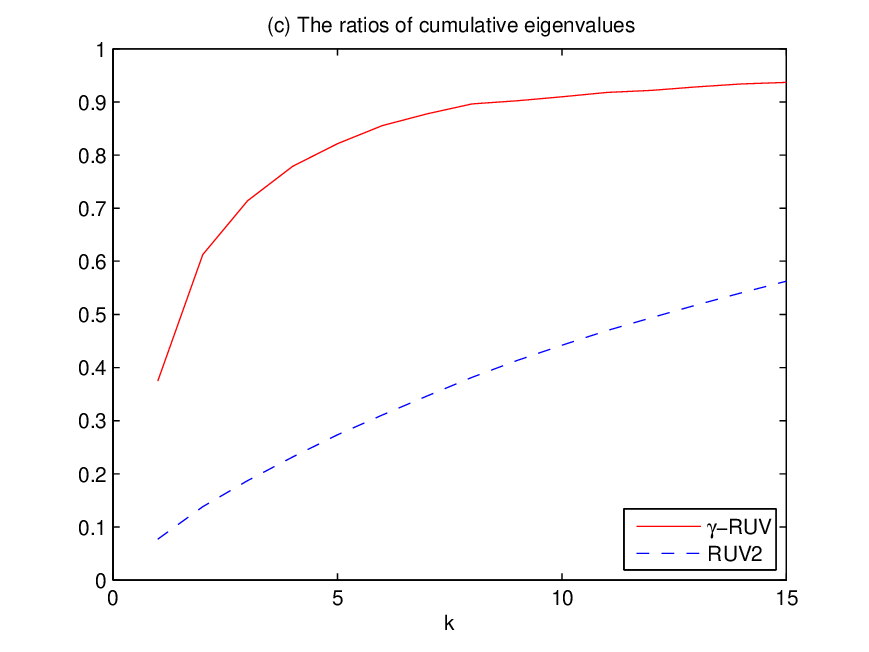}
\includegraphics[width=3in]{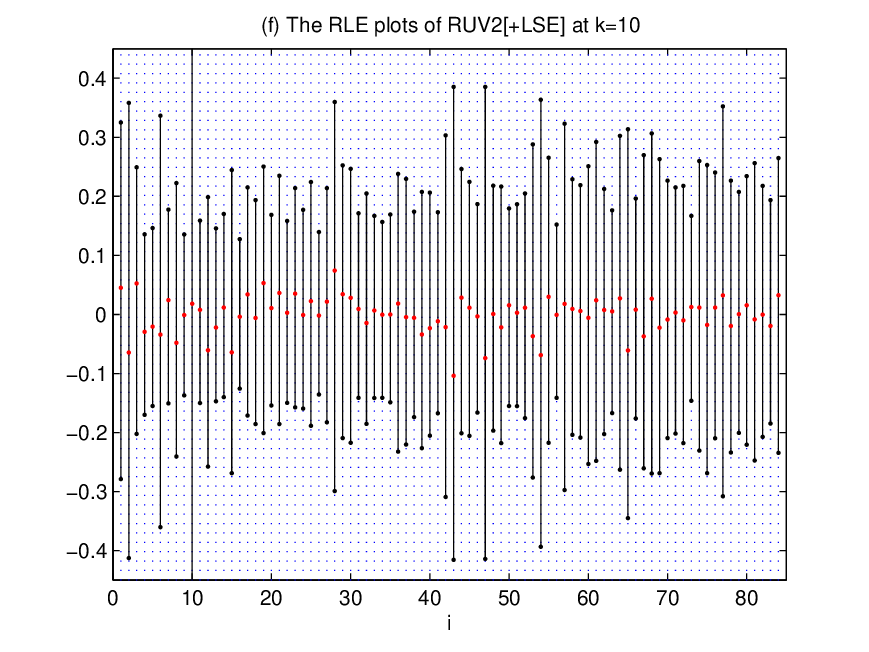}
\caption{The analysis results of the GS data with artificially added outlier matrix $\Ob$. (a) The TP of the top ranking genes at $k=10$. (b) The TP of the top 100 genes at different $k$. (c) The ratios of cumulative eigenvalues at different $k$. (d) The RLE plots of $\gamma$-RUV[+$\gamma$-LSE] at $k=10$. (e) The RLE plots of $\gamma$-RUV[+LSE] at $k=10$. (f) The RLE plots of RUV2[+LSE] at $k=10$.} \label{fig.GS_outlier}
\end{figure}

To more clearly demonstrate the power of our proposal, we report the
analysis results of the GS data with artificially added outliers. Specifically, we
generate the outlier matrix $\Ob$ as in (\ref{ol}), except
$\Wb_{(1)}$ is the batch effect taken from the GS data, and the
new analysis results with $(\pi_{\rm o},\sigma_{\rm o})=(0.05,20)$
are placed in Figure~\ref{fig.GS_outlier}. Comparing with the
results in Figure~\ref{fig.GS}, $\gamma$-RUV[+$\gamma$-LSE] can
produce roughly the same results with the original analysis. However,
the non-robust $\gamma$-RUV[+LSE] and RUV2[+LSE] are heavily affected by $\Ob$, and their TP
values are largely decreased (see Figure~\ref{fig.GS_outlier} (a)). From the RLE plots in
Figure~\ref{fig.GS_outlier}~(d)-(f) and the mean IQR values in Table~\ref{tab.lre}, it
becomes clear that $\gamma$-RUV[+$\gamma$-LSE] is the best performer
in removing unwanted variation, and achieves the smallest mean IQR value 0.186. On the other hand, the RLE plots
reveal the non-robustness RUV2[+LSE], where a portion of subjects
tend to have non-zero centers or IQR values larger than 0.2 (with
the mean IQR value $0.440$). From Figure~\ref{fig.GS_outlier}~(c),
the ratios of cumulative eigenvalues still support the
selection of $k=10$ for $\gamma$-RUV, and a high TP value is
achieved by $\gamma$-RUV[+$\gamma$-LSE] at $k=10$ (see Figure~\ref{fig.GS_outlier} (b)).
However, the ratios of
cumulative eigenvalues of RUV2 no longer provide appropriate choice
of $k$, and a low TP value is detected even with
increasing $k$ (see Figure~\ref{fig.GS_outlier} (b)). Our extra analysis confirms the robustness of
$\gamma$-RUV[+$\gamma$-LSE] in real data.

~\

\begin{center}
{\Large \textbf{References}}
\end{center}

\begin{description}
\item
Bolstad, B. M., Collin, F., Brettschneider, J., Simpson, K., Cope,
L., Irizarry, R. A., and Speed, T. P. (2005). Quality assessment of
Affymetrix GeneChip data. \emph{Bioinformatics and computational
biology solutions using R and bioconductor}, 33-47. Springer New
York.

\item
Friguet, C., Kloareg, M., and Causeur, D. (2009). A factor model approach to multiple testing under dependence. \emph{Journal of the American Statistical Association}, 104(488), 1406-1415.

\item
Fujisawa, H. and Eguchi, S. (2008). Robust parameter estimation with
a small bias against heavy contamination, {\it Journal of
Multivariate Analysis}, 99, 2053-2081.

\item
Gagnon-Bartsch, J. A., and Speed, T. P. (2012). Using control genes
to correct for unwanted variation in microarray data.
\emph{Biostatistics}, 13, 539-552.

\item
Gagnon-Bartsch, J., Jacob, L., and Speed, T. P. (2013). Removing
unwanted variation from high dimensional data with negative
controls. Technical Report 820, Department of Statistics, University
of California, Berkeley.

\item
Gandolfo, L. C., and Speed, T. P. (2017). RLE Plots: Visualising Unwanted Variation in High Dimensional Data. arXiv:1704.03590.

\item
Gerard, D. and Stephens, M. (2017). Unifying and Generalizing
Methods for Removing Unwanted Variation Based on Negative Controls.
arXiv:1705.08393.

\item
Henderson, H. V. and Searle, S. R. (1979). Vec and vech operators
for matrices, with some uses in Jacobians and multivariate
statistics. \emph{Canadian Journal of Statistics}, 7, 65-81.

\item
Hung, H., Jou, Z. Y., and Huang, S. Y. (2017). Robust mislabel
logistic regression without modeling mislabel probabilities.
\emph{Biometrics}. (DOI:10.1111/biom.12726)

\item
Kanamori, T. and Fujisawa, H. (2015). Robust estimation under heavy
contamination using unnormalized models. {\it Biometrika}, 102,
559-572.

\item
Leek, J. T. and Storey, J. D. (2007). Capturing heterogeneity in
gene expression studies by surrogate variable analysis. PLoS
genetics, 3(9), e161.

\item
Leek, J. T. (2014). svaseq: removing batch effects and other
unwanted noise from sequencing data. Nucleic acids research, 42(21),
e161-e161.

\item
Lippa, K. A., Duewer, D. L., Salit, M. L., Game, L., and Causton, H. C. (2010). Exploring the use of internal and external controls for assessing microarray technical performance. \emph{BMC Research Notes}, 3(1), 349.

\item
Price, A. L., Patterson, N. J., Plenge, R. M., Weinblatt, M. E., Shadick, N. A., and Reich, D. (2006). Principal components analysis corrects for stratification in genome-wide association studies. \emph{Nature genetics}, 38(8), 904.

\item
Risso, D., Ngai, J., Speed, T. P., and Dudoit, S. (2014). Normalization of RNA-seq data using factor analysis of control genes or samples. \emph{Nature biotechnology}, 32(9), 896-902.

\item
Sun, Y., Zhang, N. R., and Owen, A. B. (2012). Multiple hypothesis testing adjusted for latent variables, with an application to the AGEMAP gene expression data. \emph{The Annals of Applied Statistics}, 6(4), 1664-1688.

\item
Teschendorff, A. E., Zhuang, J., and Widschwendter, M. (2011).
Independent surrogate variable analysis to deconvolve confounding
factors in large-scale microarray profiling studies, {\it
Bioinformatics}, 27, 1496-1505.

\end{description}

\newpage

\newpage

\newpage

\clearpage

\clearpage

\begin{table}
\centering \caption{The mean IQR values (standard deviations) of ($\Yb-\widehat\Wb\widehat\alphab$) in the GS data. The first row represents the results of the original data, and the second row represents the results with artificially added outlier matrix $\Ob$.}\label{tab.lre} \vspace{3ex}
\begin{tabular}{cccc}
\hline
    &   $\gamma$-RUV[+$\gamma$-LSE]     &   $\gamma$-RUV[+LSE]      &   RUV2[+LSE]      \\
\hline
Without $\Ob$   &   0.172   (0.028) &   0.178   (0.029) &   0.183   (0.031) \\
Adding $\Ob$    &   0.186   (0.031) &   0.244   (0.039) &   0.440   (0.148) \\
\hline
\end{tabular}
\end{table}

\end{document}